\documentstyle{article}
\def\bigstrut{\vrule width0pt height0.6truecm}
\begin{document}
\small
\vglue -1.5truecm
\noindent
\centerline{\Large MAPPING CLUSTER MASS DISTRIBUTIONS }
\centerline{\Large VIA GRAVITATIONAL LENSING OF BACKGROUND GALAXIES}
  \centerline{ }
  \centerline{ }
 \centerline{\large T.J. BROADHURST}
 \centerline{Royal Observatory,
\bigstrut Blackford Hill, Edinburgh EH9 3HJ, UK }
 \centerline{Physics and Astronomy Dept.,
Johns Hopkins University, Baltimore,
Maryland, MD 21218, USA}
  \centerline{ }
 \centerline{\large A.N. TAYLOR}
 \centerline{Department of Astronomy, University of Edinburgh,
Blackford Hill, Edinburgh EH9 3HJ, UK}
  \centerline{ }
 \centerline{\rm AND}
  \centerline{ }
 \centerline{\large J.A. PEACOCK}
 \centerline{Royal Observatory, \bigstrut Blackford Hill,
Edinburgh EH9 3HJ, UK }
%
%
\def\ref{\parskip=0pt\par\noindent\hangindent\parindent
    \parskip =2ex plus .5ex minus .1ex}
\def\gs{\mathrel{\raise1.16pt\hbox{$>$}\kern-7.0pt
\lower3.06pt\hbox{{$\scriptstyle \sim$}}}}
\def\ls{\mathrel{\raise1.16pt\hbox{$<$}\kern-7.0pt
\lower3.06pt\hbox{{$\scriptstyle \sim$}}}}
\def\pmb#1{\setbox0=\hbox{#1}%
  \kern-.025em\copy0\kern-\wd0
  \kern.05em\copy0\kern-\wd0
  \kern-.025em\raise.0433em\box0 }
\def\del{\pmb{$\nabla$}}
\def\ss{\rm\scriptscriptstyle}
\def\half{\textstyle{\scriptstyle 1\over\scriptstyle 2}}
\def\Var{{\rm Var}}
%
%
\newcommand{\btheta}{\mbox{\boldmath $\theta$}}
\newcommand{\balpha}{\mbox{\boldmath $\alpha$}}
\newcommand{\be}{\begin{equation}}
\newcommand{\ee}{\end{equation}}
\newcommand{\ba}{\begin{eqnarray}}
\newcommand{\ea}{\end{eqnarray}}
\newcommand{\bfig}{\begin{figure}}
\newcommand{\efig}{\end{figure}}
\newcommand{\et}{{\em et al. }}
\newcommand{\ie}{{\em ie. }}
\newcommand{\rgl}{\rangle}
\newcommand{\lgl}{\langle}
\newcommand{\x}{{\bf x}}
\newcommand{\z}{{\bf z}}
\newcommand{\k}{{\bf k}}
\newcommand{\y}{{\bf y}}
\newcommand{\p}{{\bf p}}
\newcommand{\q}{{\bf q}}
\newcommand{\g}{{\bf g}}
\newcommand{\r}{{\bf r}}
\newcommand{\s}{{\bf s}}
\newcommand{\tbf}{{\bf t}}
\newcommand{\n}{{\bf n}}
\newcommand{\hx}{\hat{x}}
\newcommand{\hs}{\hat{s}}
\newcommand{\htt}{\hat{t}}
\newcommand{\hxb}{\hat{\x}}
\newcommand{\vb}{{\bf v}}
\newcommand{\pot}{\varphi}
\newcommand{\txb}{\tilde{\x}}
\newcommand{\tx}{\tilde{x}}
\newcommand{\tJ}{\tilde{J}}
\newcommand{\tP}{\tilde{P}}
\newcommand{\tmu}{\tilde{\mu}}
\newcommand{\cJ}{{\cal J}}
\newcommand{\cT}{{\cal T}}
\newcommand{\cI}{{\cal I}}
\newcommand{\cW}{{\cal W}}
\newcommand{\al}{\alpha}
\newcommand{\bet}{\beta}
\newcommand{\txi}{\tilde{\xi}}
\newcommand{\trho}{\tilde{\rho}}
\newcommand{\tdelta}{\tilde{\delta}}
\newcommand{\tDelta}{\tilde{\Delta}}
\newcommand{\lvcv}{\varepsilon}
\newcommand{\tb}{\tau}
\newcommand{\xt}{\x,t}
\newcommand{\zz}{\q,\tb}
\newcommand{\xx}{\x(\q,\tb),\tb}
\newcommand{\txx}{\txb(\q,\tb),\tb}
\newcommand{\de}{\partial}
\newcommand{\daa}{\frac{\dot{a}}{a}}
\newcommand{\idaa}{\frac{a}{\dot{a}}}
\newcommand{\dxj}{\frac{\de}{\de x^j}}
\newcommand{\dxi}{\frac{\de}{\de x^i}}
\newcommand{\dxai}{\frac{\de}{\de x_1^i}}
\newcommand{\dxbj}{\frac{\de}{\de x_2^j}}
\newcommand{\dxk}{\frac{\de}{\de x^k}}
\newcommand{\dxm}{\frac{\de}{\de x^m}}
\newcommand{\dxp}{\frac{\de}{\de x^p}}
\newcommand{\dqj}{\frac{\de}{\de q^j}}
\newcommand{\dqi}{\frac{\de}{\de q^i}}
\newcommand{\dkj}{\frac{\de}{\de k^j}}
\newcommand{\dki}{\frac{\de}{\de k^i}}
\newcommand{\dt}{\frac{d}{d \tb}}
\newcommand{\xib}{\mbox{\boldmath $\xi$}}
\newcommand{\lvcvb}{\mbox{\boldmath $\lvcv$}}
\newcommand{\omegab}{\mbox{\boldmath $\omega$}}
\newcommand{\nab}{\mbox{\boldmath $\nabla$}}
\newcommand{\aln}{& &\!\!\!\!\!}
\newcommand{\e}{\equiv}
\newcommand{\ssim}{\approx}
\newcommand{\nn}{\nonumber \\}
\newcommand{\lfe}{\lefteqn}
\newcommand{\fig}{\centerline{{\bf Figure}}}
\newcommand{\nd}{\noindent}
\newcommand{\indi}{\hspace{2in}}
\newcommand{\nwl}{\newline}
\newcommand{\nds}{\noindent \hspace*{0.5 mm}}
\newcommand{\spss}{\hspace{1 cm}}
\newcommand{\apj}{ApJ}
\newcommand{\astr}{Astron. J.}
\newcommand{\apl}{Astrophysical Lett.}
\newcommand{\mn}{MNRAS}
\newcommand{\na}{Nature}
\newcommand{\physsc}{{\sl Phys. Scripta.}}
\newcommand{\asap}{{\sl Astr. Astrophys.}}
\newtheorem{figudumx}{Figure}
\newenvironment{figdumy}{\begin{figudumx}\rm}{\end{figudumx}}
\newcommand{\bfg}{\begin{figdumy}}
\newcommand{\efg}{\end{figdumy}}

\centerline{}
\begin{abstract}
\noindent
We present a new method for measuring the projected mass distributions
of galaxy clusters, based solely on the gravitational lens
amplification of background galaxies by the cluster potential field.
The gravitational amplification is measured by comparing the joint
distribution in redshift and magnitude of galaxies behind the cluster
with that of the average distribution of field galaxies. Lensing
shifts the magnitude distribution in a characteristic
redshift--dependent way, and simultaneously dilutes the surface
density of galaxies. These effects oppose, with the latter dominating
at low redshift and the former at high redshift, owing to the
curvature of the galaxy luminosity function.
Lensing by a foreground cluster thus induces an excess of bright
high-redshift galaxies, from which the lens amplification
may be inferred.

\hglue 1.0truecm
 We show that the total amplification is directly related to the
surface mass density in the weak field limit, and so it is possible to
map the mass distribution of the cluster. The method is shown to be
limited by discreteness noise and galaxy clustering behind the lens.
Galaxy clustering sets a lower limit to the error along the redshift
direction, but a clustering independent lensing signature may be
obtained from the magnitude distribution at fixed redshift. Provided
the luminosity function deviates from a pure power law, the
lens--induced brightening can be measured directly by comparison with
the field.  In the limit that galaxy luminosities are independent of
environment, this method is only shot--noise limited.

\hglue 1.0truecm
 Statistical techniques are developed for estimating the surface mass density
of the cluster. We extend these methods to account for any obscuration by
cluster halo dust, which may be mapped
independently of the dark matter.  We apply the method to a series of numerical
simulations and show the the feasibility of the approach.  We consider the
use of approximate redshift information, and show how the mass
estimates are degraded; finally we discuss the data required
to map the dark matter in clusters from photometry alone.

\hfill\break

\end{abstract}

\section{Introduction}

Quantifying the distribution of dark matter in the
Universe is one of the most important challenges for modern
cosmology. Galactic rotation and the dynamical behaviour of groups of
galaxies have long shown the presence of mass far in excess of that
due to normal stellar populations. As the
characteristic scale of structure increases, it appears that
dynamical--to--luminous mass ratios increase,
consistent with a scenario in which luminous matter is biased
towards high density environments.  On the largest scales, $30$ to $100 \,
h^{-1}$Mpc, the situation favoured by large-scale flows and
perturbations in the galaxy number density is one in which mass
roughly traces light and the mean mass density is near critical
(e.g. Dekel 1994)

On intermediate scales, rich clusters are of special interest as they
represent the largest gravitationally bound objects. Measurement of
cluster mass profiles would provide information on the formation
history of structure -- a problem inexorably bound up with the thermal
properties of the dark matter and the cosmological model.
The line-of-sight velocity dispersions of galaxies,
plus thermal X-ray data, indicate a density parameter
of $\Omega_0\simeq 0.1$ if the $M/L$ ratios from the
central 1 -- 2 $h^{-1}$Mpc of clusters are representative.
However, these observations are limited in two respects.
Firstly, both estimates rely on the assumption of equilibrium models
to relate observables to the mass distribution. Substructure in the
density or velocity fields is neglected, as are perturbations due
to accretion, a simplification which is no longer supported by detailed
observations (White, Briel \& Henry 1993). Secondly, as both methods rely on
galaxies or gas to act as tracers, they are restricted to
the central core of the cluster.
In order for $\Omega=1$ to be tenable, we either require existing
central mass estimates for clusters to be a severe underestimate,
or that more mass is lurking at larger radii. In all bias models
that have been discussed, this would take the form of
an isothermal halo extending out to roughly $10 \, h^{-1}$ Mpc.

It is clearly of great importance to test these alternatives,
and a new powerful method has been to use the properties of
gravitational lensing of images behind the cluster to probe the
integrated mass profile.  The shear distortion of images by
gravitational lensing has been explored in detail since the pioneering
work of Tyson {\it et al.} (1984). The statistical distortion of
off-axis images behind extended mass distributions has been well
explored theoretically ({\it e.g.} Kochanek 1990, Miralda--Escud\'e
1991, Kaiser \& Squires 1993) and there is widespread interest in
using observational data to detect and interpret this effect (Tyson
\et 1990; Kneib \et 1994; Smail \et 1993, 1994;
Fahlman \et 1994; Bonnet, Mellier \& Fort 1994).
Conclusive detections are difficult due to the requirement
of obtaining data of high enough quality that higher order
moments of faint galaxy images can be measured in the presence of
atmospheric seeing and the fundamental limitation imposed by the
intrinsic dispersion of galaxy ellipticities. Also, the forms of
cluster mass distributions are not known in advance so that ambiguity
arises in choosing appropriate potentials to test against the data.

This latter problem has been overcome, in the limit of weak
amplifications, by Kaiser \& Squires (1993). They demonstrate a
nonlocal reconstruction technique for deriving an arbitrary projected
potential based on the distortion of galaxy ellipticities.
Initial applications of this method by Fahlman \et 1994, Smail
\et 1994 and Bonnet \et 1994 have yielded very encouraging
detections of dark matter in the central $\sim 1$ Mpc of several clusters.
However, this approach is still not completely general.
In the case of sheet--like matter distributions, images
are not distorted, and hence the potential cannot be recovered.
By working only with the shear, the first-order effect
of the lens (its amplification) is
ignored. Although large numbers of galaxies allow the higher order
effects of derivatives of the mass distribution to be detected in
principle, it would clearly be preferable to have a more direct
method.

In this paper we investigate an alternative approach to recovering the
surface potential of clusters which relies solely on the
amplification of galaxies in the cluster background.  Our method is
simply to compare the bivariate distribution of redshift and
magnitude, $N(m,z)$, of galaxies behind the cluster with that of the
general field, to the same magnitude limit.  The observable difference
arises solely from the amplification of distant galaxy images. We show
that in the weak field limit this amplification can be related
directly to the surface mass density. This allows us to use the
transformation of $N(m,z)$ with radius from the center of the cluster as
a model independent estimate of the mass distribution of the cluster.

The method is of course limited by the shot noise due to finite
numbers of background galaxies.
Fluctuations in the redshift and magnitude
distribution due to the intrinsic clustering of galaxies in the
background are also a potential problem.
However, these only affect the redshift
distribution; the magnitude distribution at each redshift can be used
to give a mass estimate which may be compared with the independent
measurement obtained from distortion in the redshift distribution.

In general, the methods we present here can be thought of as a natural
extension of the faint imaging work by the addition of redshift information for
the faint galaxies. Imaging at high spatial resolution can constrain the shape
of the potential with a smoothing scale of 0.5 -- 1 arcmin. However,
we argue that the depth of the potential, and therefore the general M/L, is
most reliably constrained by full redshift-magnitude information.
The angular resolution that our method can attain is coarser by a
factor of a few than what is possible with imaging
(in a given area of sky there are many fewer galaxies
to a practical spectroscopic limit to average over, but the signal
is stronger). The alternatives of image shear and redshift
information are thus complementary, allowing
the fullest characterisation of the cluster potential.

The paper is organized as follows. In Section 2 we set
out the basic principles of the lensing distortion effect on the
redshift--magnitude distribution. The lens amplification and the
background population of galaxies are also discussed.  In Section 3 we
show that in the regime of weak gravitational lensing, the distortion
can be related to the surface potential of the lensing cluster for any
reasonably smooth, but otherwise arbitrary potential distribution.
The effects of the background cosmological model on the distortion and
possible confusion with extinction by cluster dust is also considered.
Maximum likelihood reconstruction techniques are presented in Section
4, where we consider the effects of intrinsic galaxy clustering and
discreteness.
Section 5 discusses the application of these methods in practice.
We present a number of numerical realizations of the
effect and apply the likelihood estimators to these.
We consider observational strategies to minimize the amount
of telescope time needed to detect dark matter, either in
an individual cluster or statistically by averaging over clusters.
Section 6
considers the price paid in accuracy and bias of the results if
redshift estimates of restricted accuracy are used. Finally, in
Section 7, we summarize our conclusions.

\section{Lensing of background galaxies}

The general distribution of galaxies in redshift and
luminosity is described by a bivariate luminosity function,
$\phi(L,z)$, which gives the comoving number density of galaxies per
unit interval in redshift and luminosity.  Allowing for galaxy
clustering, this can be decomposed into
\be
\phi(L,z)  = \phi_0(L,z)  [ 1+ \delta({\bf x}, L)],
\label{e:1}
\ee
where $\phi_0$ is the (evolving) expectation of the luminosity
function.  In the limit that galaxy luminosity is independent of
environment (assumed hereafter), the density perturbation $\delta$ is
independent of $L$.

The relation of this luminosity function to the observed bivariate
distribution $N(m,z)$ is just
\be
N(m,z) \; dm\; dz=\phi(L,z)\; dL\; dV(z).
\ee
Here $N(m,z)$ is the number of objects per steradian in
the interval $dm\; dz$. The luminosity $L(z)$ is given by
\be
L(z) = 4 \pi\, S\, d_{\ss L}^{2}(z) (1+z)^{\alpha-1},
\ee
where $d_{\ss L}(z)$ is the luminosity distance to the source, $S$ is
the flux density of the source, and $\alpha$ is the effective spectral
index of the source (in the sense $S\propto \nu^{-\alpha}$); note that
we do not assume power-law spectra -- this is just another way of
writing the K-correction.

What is the effect of a gravitational lens on a  background population of
galaxies? Lensing is well-known to conserve surface
brightness, distorting image shapes and increasing
their area by some amplification factor $A$.
The effect this has on a galaxy catalogue created from lensed
data depends on how galaxy magnitudes are defined.
Most commonly these are isophotal, in which case the
effect of lensing is to scale galaxy fluxes in the
same way as for a point source: increase by a factor $A$.
For fixed angular apertures things are more complex, however.
If the growth curve for galaxy flux is a power law in
radius, $L(<r)\propto r^\epsilon$, then lensing increases
apparent fluxes by $A^{1-\epsilon/2}$. Since realistic
values of $\epsilon$ are $\simeq 0.5$, this is not very
different from the behaviour of isophotal magnitudes;
the use of apertures thus weakens the sensitivity of the
methods discussed here by a factor $\simeq 0.75$. In what
follows, we shall assume isophotal magnitudes, and neglect
this factor.

The effect of the foreground lens is then
simply to translate the apparent magnitude
distribution of background galaxies
by some redshift-dependent amplification
factor, $A(z)$, and to reduce the surface
density through the simultaneous distortion of
the angular distances between galaxies:
\be
N'(m, z)= N(m+2.5\log_{10} A(z),z)\, / \, A(z).
\ee
We show below that the function $A(z)$ will usually depend on only one
parameter: the surface density of the lens.  Given knowledge of the
unperturbed counts, it would then be tempting to use maximum
likelihood and the above equation to determine the lens density by
fitting to the 2-dimensional $(m,z)$ distribution.  However, this is
not so straightforward because the numbers of galaxies in a given
redshift bin are subject to fluctuations caused by galaxy clustering.
In contrast, the distribution in magnitude can be renormalized
at each redshift slice to eliminate clustering variations.  It
therefore makes sense to consider extracting information from the
redshift and magnitude axes separately.

\subsection{Lensing and the Redshift Distribution}

Suppose we integrate over apparent magnitude to obtain $n(z)$ --
the number of galaxies in the redshift interval $dz$  per steradian:
\be
n(z) \;dz =  dV(z) \;\int_{L_{\rm min}}^\infty \phi(L,z) \;dL,
\ee
with intrinsic luminosity
above the limiting luminosity set by the flux limit of the detector.
The effect of placing a lens in front of a magnitude limited  galaxy
distribution is to decrease the effective flux limit,
due to conservation of surface brightness.
{}From equation (4), we see that  the effect of lensing on the redshift
distribution is
\ba
n'(z) dz &=& A^{-1}\, \Phi[L_{\rm min}(z)/A, z] \, dV(z), \nn
         &\simeq& A^{\beta(z)-1} n(z) \, dz,
\label{eq: e2.1}
\ea
where the integral luminosity function is denoted by $\Phi$.
Here we have approximated the luminosity function as a
power law, where
\be
\beta(z) \equiv -\; \frac{d \ln \Phi[L(z),z] }{ d \ln L(z) }
\ee
is the effective index of the luminosity function.  As far as the
redshift-distribution distortions are concerned, this is a quite
accurate enough approximation and considerably simplifies the
analysis.  From this definition of $\beta(z)$ and $L(z)$, we see that
at small redshift the angular scattering of images by the lens will
dominate, and the redshift distribution will drop.  Meanwhile at large
redshift the increase in the total number of observable galaxies will
dominate. In Figure 1 we show this distortion for a range of
amplification factors. As expected, there is a node where the effects
cancel. Hence, given a suitable model for the unlensed distribution
function's power-law index (see Section 2.4) the change in the shape
of $n(z)$ with redshift can be measured, thus yielding $A(z)$.
Although this increasing tail of high-redshift objects is clear enough
in theory, the accuracy to which the shape of $n(z)$ can be determined
is limited by galaxy clustering fluctuations and shot noise.  We
consider these in quantitative detail in Section 4.2.

\subsection{Lensing and the Luminosity Distribution}

Even in the presence of very large and unknown density fluctuations as
a function of redshift, it is still possible to detect a lensing
signal.  The amplification of galaxy luminosities due to the lens
results in a shift in the luminosity function, which can be studied in
a clustering--independent manner if the luminosity function is
renormalized to become the probability distribution, $P(m|z)$.  In the
limiting case of a pure power law luminosity function, $\Phi(L)
\propto L^{\beta}$, this probability distribution is invariant to any
shift, and no distortion can be measured.  However, realistic
distributions in luminosity have a non--power law cutoff.  As a
example, consider an exponential luminosity function
$\Phi=\exp(-L/L_*)$. It is a simple exercise to use the normalized
version of this distribution to obtain the maximum-likelihood estimate
of $L_*$ and its uncertainty from a set of $n$ galaxies (see Section
4). For an assumed $L_*$, this gives an estimate of the amplification,
with a fractional rms accuracy
\be
{\sigma_A\over A}={1\over \sqrt{n}}.
\ee
Although it is possible to find pathological examples such as the top
hat distribution where the existence of a sharp feature means that the
fractional error in $A$ goes as $1/n$, any distributions of practical
interest will always obey the $1/\sqrt{n}$ scaling -- although the
coefficient may be $\gg 1$ if the distribution is close to a power
law.

These methods provide us with two independent methods for estimating
the strength of the lens. One results from the net effect of the
amplification and area dilution in redshift, the second is a redshift
dependent shift in the magnitude distribution from the amplification
alone. Each method provides an independent measure of the lens
amplification, and has the advantage of restricting the effects
of galaxy clustering to only one of the methods.

\subsection{Lens Amplification}

The amplification factor, $A(z)$, is the determinant of the
deformation tensor describing the mapping from source plane to the
image plane (see Section 3), and can in general be expressed in terms
of the convergence and shear distortions of an object in the source
plane onto the image plane (Young 1981; Miralda--Escud\'e 1991). Thus
the amplification can be expressed as the increase in total surface
area,
\be
A = \frac{1}{(1-\kappa)^2-\gamma^2},
\ee
where $\kappa$ is the amplitude of the convergence and $\gamma$ the
amplitude of the shear of the image.  For example, in the case of a
singular isothermal cluster, $\gamma =
\kappa= \Sigma/\Sigma_{\ss C}$, where $\Sigma_{\ss C}$ is the critical
surface density producing a caustic for a sheet lens:
\be
\Sigma_{\ss C}(z) \equiv
  \frac{c^2\;D_{\ss S}}{4 \pi G D_{\ss L} D_{\ss LS}}.
\ee
$D_{\ss L}$ and $D_{\ss S}$ are the angular distances to the lens and
source, and $D_{\ss LS}$ is the angular distance of the source
as seen at the lens (we shall use the filled--beam approximation).
In the specific case of $\Omega_0=1$, the angular distance is
\be
D_{\ss LS}(z_{\ss L},z_{\ss S})=
{2c\over H_0}\,{[(1+z_{\ss L})^{-1/2}-
(1+z_{\ss S})^{-1/2}] \over (1+z_{\ss S})},
\ee
where $z_{\ss L}$, and $z_{\ss S}$ are the redshifts of the lens and
source, respectively. In Section 3.2 we shall discuss the effects of
altering the cosmological model.

We have now assembled the necessary expressions for reconstructing the
surface density of clusters.  Given the lens-distorted $N'(m,z)$ and a
form for the true $N(m,z)$, we can calculate the amplification factor.
In the next Section we shall discuss what is known empirically of the
background population, while in Section 3 we show under what general
conditions the amplification factor can be related to the surface
density.

\subsection{The background galaxy population}

The redshift distribution of faint field galaxies, and its magnitude
dependence, is now being defined directly by redshift surveys ({\it
e.g.} Colless {\it et al.} 1993; Glazebrook {\it et al.} 1994).  We
shall use results from a model designed to fit these data by making
specific assumptions about cosmology and evolution of the luminosity
function.  However, it is important to emphasize that the functions we
require [$N(m,z)$, $n(z)$, $\beta(z)$, defined in equations 2, 5 and
7] are in principle directly observable and model-independent.

One possibility would be to use the model discussed by Broadhurst,
Ellis \& Glazebrook (1992).  This assumes $\Omega=1$ and a galaxy
population divided into five distinct types from irregular to
elliptical, together with appropriate K-corrections. The luminosity
function undergoes `merging' evolution in which galaxies are typically
less luminous but more numerous in the past; this yields the required
excess numbers of faint galaxies without producing a large number of
(unobserved) galaxies at $z\gs 1$.  It would be possible to work
directly with the population of the $(m,z)$ grid output by such a
model.  However, since this model is relatively complex, we have
chosen to illustrate the results of this paper in terms of a simpler
analytic construction.
We use a single Schechter function which undergoes
a combination of luminosity and density evolution:
\be
\phi(L) =\phi^*(z)\, \exp \left[ -  L/L_* \right],
\ee
where $\phi^*(z)=0.02h^3(1+z)^2\;\rm h^{-1}Mpc^{-3}$, and luminosity
evolution is simulated by assuming a constant $L_*$, and a constant
spectral index $\alpha=3$ (i.e. $K(z)=5\log_{10}[1+z]$);
For practical comparisons, we choose to work in the
$R$ band, in which the value $M^*_R=-21.5$ for $h=1$
is appropriate for this model.
$\Omega=1$ is assumed, but all that is needed is the empirical
$m$ -- $z$ distribution, which is independent of cosmological assumptions.
This model is in fact quite realistic: it gives a good fit to
the observed $R$-band counts (Metcalf \et 1994), and predicts
median redshifts in accord with observation to the limit of existing data.

Given the luminosity function, it is easy to find the numerical
results for $n(z)$, and $\beta(z)$.
For practical purposes, we will often be interested in working
at the faint limit for spectroscopy, which we take to be $R=22.5$.
The $n(z)$ and $\beta(z)$ functions for this case
may be fitted directly by
the following expressions, which we have adopted for convenience at
the appropriate points of the analysis:
\be
n(z)= 11.7\; z^{1.63}\, \exp[-(z/0.51)^{1.79}],
\ee
\be
\beta(z)=0.15+0.6z+ 1.1z^{3.2}.
\ee
Recall that $n(z)$ refers to the probability distribution for redshift
and may be normalised to the cumulative surface density (which at
$R=22.5$ is approximately 20,000 per square degree).

\section{Weak lensing limit}

\subsection{Smooth lensing potentials}

We have shown above that, in principle, the mean amplification
over part of a cluster can be produced by obtaining data for
background galaxies. The question now is what information this would give
us about the mass distribution. In fact, under certain very reasonable
assumptions, it turns out that one is able to recover the projected
mass distribution of the cluster directly.

Consider the case of sources at a given redshift, so that the lensing
equation can be written in terms of the lensing potential:
\be
{\pmb{$\theta$}}={\pmb{$\phi$}}+ \del\psi.
\ee
The amplification of background images is just the reciprocal of the
Jacobian determinant, given by
\be
A^{-1}=\left|{\rm det}\left(\delta_{ij}-{\partial^2\psi\over
\partial\theta_i\partial\theta_j}\right)\right|.
\ee
Except in the case of strong lensing, the determinant is always
positive and the amplification becomes
\be
A^{-1}=1-\nabla^2\psi+{\partial^2\psi\over\partial
x^2}{\partial^2\psi\over\partial y^2} -
\left[{\partial^2\psi\over\partial x\partial y}\right]^2
\ee
(defining 2D angular cartesian coordinates $x$ \& $y$:
$\theta=\sqrt{x^2+y^2}$).
Since Poisson's equation in this context says
\be
\nabla^2\psi=2{\Sigma\over\Sigma_{\ss C}},
\ee
the surface density of the lens may be measured directly if the terms
nonlinear in potential derivatives may be dropped.  This would be the
case if the lens was simply a screen of constant surface density; the
question is therefore to what extent this is a reasonable
approximation to the mass distribution in the outer parts of clusters.

Although we will almost always be working in the weak lensing regime
$A-1\ll 1$, this alone does not guarantee that the shear terms are
negligible: any given amplification can be achieved with zero surface
density, given an appropriate degree of shear. To make progress, we
need in addition to assume that the lens is {\it smooth\/}.  This is a
reasonable assumption in the case of cluster dark matter, where there
is a characteristic angle in the form of the Einstein-ring radius
$\theta_{\ss E}$.  We know from the lack of multiply-imaged galaxies
in the outer parts of clusters that the dark matter does not contain
structure on the arcsecond scale (apart from individual cluster
galaxies, which must be allowed for separately).  A general lens may
have its lensing potential described as a 2D Fourier transform:
$\psi=\sum \psi_k \exp -i{\bf k\cdot r}$.  Constructing the second
derivatives of this expression and squaring, we see that all
derivatives will have similar mean square values if the potential
fluctuations are reasonably isotropic
\be
\langle \psi_{xx}^2 \rangle \sim
\langle \psi_{yy}^2 \rangle \sim
\langle \psi_{xy}^2 \rangle \sim
\sum |\psi_k|^2 k^4 \sim \left[{\Sigma_{*}\over \Sigma_{\ss C}}\right]^2,
\ee
where we have used Poisson's equation to define a typical surface
density, $\Sigma_{*}$, about which the lens fluctuates.  If the
potential is grossly anisotropic, so that it contains wavevectors
pointing only in one direction, the shear vanishes, so statistical
isotropy is the worst case.  We can therefore write
\be
{\Sigma\over\Sigma_{\ss C}}={1-A^{-1}\over 2} +
  O\left(\left[{\Sigma_{*}\over \Sigma_{\ss C}}\right]^2\right).
\ee
Since it is reasonable to assume that the typical surface density
declines roughly monotonically with radius in a cluster, it should
therefore be a good approximation in practice to neglect shear and
deduce the surface density directly from amplification, provided only
that the amplification factor is close to unity.  This procedure must
break down sufficiently near the center, but at this point there
are in any case few background galaxies over which to average;
it is better to constrain the central projected mass by using
arcs from individual galaxies near caustics.

\subsection{Effects of cosmological model in the weak field limit}

As we have shown in Section 2, the distortion effect of the lens is
completely characterized by the amplification function $A(z)$. It is
convenient to write this, in the weak field approximation as
\be
A(z) = 1 + 2 \kappa(z),
\ee
where $\kappa$ is the amplitude of the convergence (Section 2.3).
As this is a function of the critical surface density, defined by the
equation
\be
\kappa(z) = \frac{\Sigma}{\Sigma_{\ss C}(z)},
\ee
which is itself a function of angular distance, the resulting
amplification is also a function of cosmological model. We can
absorb some of the dependence on cosmological model by
parameterizing the lens via the value of $\kappa$
for a source placed at infinity: $\kappa_{\infty}$. Note that
expressing the potential in this limit is done only for
theoretical convenience: for practical calculations one must use the
smaller quantity $\kappa(z)$.

For the modelling of lensing amplification, we need to know
whether the redshift dependent quantity $\kappa/\kappa_\infty$
depends significantly on cosmological model. If the
cosmological constant is zero, we have
\be
{\kappa(z)\over \kappa_\infty} =
{ g(z_{\ss L})[2-\Omega_0+\Omega_0 z_{\ss S}] -
  g(z_{\ss S})[2-\Omega_0+\Omega_0 z_{\ss L}] \over
  g(z_{\ss L})[2-\Omega_0+\Omega_0 z_{\ss S} + (\Omega_0-2)g(z_{\ss S})] },
\ee
where $g(z)\equiv \sqrt{1+\Omega_0 z}$ (Refsdal 1966).
Figure 2a shows the ratio $\kappa(z)/\kappa_{\infty}$ for
$\Omega_0=0.1$ and $\Omega_0=1$ and for lenses at redshifts
$z_{\ss L}=0.1$ to $0.4$. For a given lens redshift these models differ
little over this range.

In the other case of interest, that of a flat model with a
nonzero cosmological constant, we must use
$D_{\ss LS}=D(z_{\ss S})-D(z_{\ss L})(1+z_{\ss L})/(1+z_{\ss S})$ and
\ba
D(z) &=&{c\over H_0}\; \frac{1}{(1+z)} \int^z_0
\frac{dz}{[(1-\Omega_0)+\Omega_0(1+z)^3]^{1/2}}, \nn
  &\simeq& {c\over H_0}\; \frac{z}{(1+z)(1+3\Omega_0z/4)}.
\ea
As no expression for the angular distance exists in closed form
(Dabrowski \& Stelmach 1986), we have given an approximate expression
in the second line by expanding the integrand to first order.  This is
sufficiently accurate for $z < 1$.  The amplification ratio,
$\kappa(z)/\kappa_{\infty}$, for this model is shown in Figure 2b for
$\Omega_0=0.1$ and $\Omega_0=1$, again for a range of lens redshifts.
 As in the open model, the effect on
the amplification is marginal.
Hence we shall only consider the
Einstein--de-Sitter model, where
\ba
\frac{\kappa(z)}{\kappa_\infty} &=& f(z), \nn
f(z) &=& {\sqrt{1+z}- \sqrt{1+z_{\ss L}} \over
\sqrt{1+z}-1}.
\ea

To relate the dimensionless surface-density measure $\kappa_\infty$
to physical values, we need the critical value of surface density
\be
\Sigma_{\ss C}=10^{15.22}M_\odot{\rm Mpc}^{-2}
{(D_{\ss S}/{\rm Gpc}) \over
(D_{\ss L}/{\rm Gpc}) (D_{\ss LS}/{\rm Gpc}) }.
\ee
Again using Refsdal's (1966) result, this gives the physical
surface density in terms of $\kappa_\infty$ as
\ba
\Sigma=10^{14.44}&&\!\!\!\!h\;M_\odot{\rm Mpc}^{-2} \;\times\nn
&&{\Omega^2 (1+z_{\ss L})^3 \over
g(z_{\ss L})[\Omega z_{\ss L}+(\Omega-2)(g(z_{\ss L})-1)]}\; \kappa_\infty
\ea
(for zero $\Lambda$).
Again, this is rather insensitive to $\Omega$, particularly since
we are usually interested in relatively low lens
redshifts, $\z_{\ss L}\ls 0.5$. Only a few \% error is
introduced by using the $\Omega=1$ form in this regime:
\be
\Sigma=10^{14.44}\,h\;M_\odot{\rm Mpc}^{-2} \;
{ (1+z_{\ss L})^2 \over \sqrt{1+z_{\ss L}} -1 }\; \kappa_\infty.
\ee
By comparison, the surface density for an isothermal sphere is
\be
\Sigma={\sigma^2\over 2Gr} = 10^{14.07} \;M_\odot{\rm Mpc}^{-2} \;
\sigma^2_{1000}\; r_{\rm Mpc}^{-1}.
\ee
As a practical example,
we might be interested in measuring  $\Sigma$ at
$1h^{-1}$ Mpc for a system with velocity dispersion
$\sigma=1000\,\rm km\,s^{-1}$, so this
corresponds to $\kappa_\infty=0.035$ at $z_{\ss L}=0.3$.
Further examples are the distortion-based measurement
of Fahlman \et (1994) on ms1224 at $z_{\ss L}=0.33$, which
converts to an average ${\bar\kappa}_\infty=0.15\pm0.04$ within a radius of
$0.48 h^{-1}$ Mpc.
Alternatively, consider Abell 370 at a redshift $z_{\ss L}=0.374$. For a source
galaxy at $z_{\ss S}=0.724$, the Einstein radius is inferred
from the principal arc curvature to be  $25''$ (Grossman
\& Narayan 1989). If the dark matter were a simple isothermal
sphere, we would have $\kappa(0.724)=0.5$ at this point (a
radius of $0.078h^{-1}$ Mpc).
At a radius of 1 $h^{-1}$Mpc corresponding to an
angular radius of about $5.4'$, we would then expect $\kappa_\infty
\simeq 0.09$.
It is therefore clear that an interesting and competitive
level of sensitivity for our method will require an rms uncertainty
of $\ls 0.05$ in $\kappa_\infty$.

\subsection{Obscuration by cluster halo dust}

In addition to lensing, dust in cluster halos will reduce the
surface density of observable galaxies. As this is a local effect we
model the inclusion of this obscuration by an additional
redshift--independent convergence term, $\kappa_{\rm dust}$ (which
will generally be negative).  In the weak field limit the total
convergence measured by an observer is
\be
\kappa(z) = f(z)\kappa_{\infty} + \kappa_{\rm dust}.
\ee
However, the dust `amplification' will not dilute numbers on the sky
by the factor $1/A(z)$, due to lensing; this and the different
redshift dependence are two ways in which dust can be distinguished
from lensing.

Limits can be placed on the abundance of dust in cluster haloes by
the cluster avoidance effect of quasars (Boyle \et 1988), although
some dilution is of course expected from lensing itself, given the
relatively flat count slope for quasars for B$>$19. More useful are
the photometry studies of cluster members (Bower \et 1992, Ferguson
1993) for which little extinction is claimed. For illustration taking
$0.^{m}5$ of extinction as a fiducial upper limit we find that
$|\kappa_{\rm dust}| \ls 0.3$. As this introduces an element of
uncertainty into the following analysis, we shall leave
$\kappa_{\rm dust}$ as a free parameter, to be fixed by the
observations themselves.

There are two further factors relating to the distribution of dust in
the model that we shall now address: the effects of intergalactic
dust, or that in intervening galaxies, and the reddening effect of
galaxy colours due to scattering by cluster dust. We shall assume
there that the intergalactic dust is negligible, since at high
redshift this would obscure quasar emission in the optical. Dust in
intervening galaxies or clusters is also assumed negligible, given
that the probability of multiple objects lying along the line of sight
is low (Press \& Gunn 1973), and hence so is further obscuration.

The effect of reddening only becomes significant in the event
that we wish to use estimates of redshift (see Section 6). In this
case a model dependent correction to the colour--magnitude relation
may have to be applied before it is used to estimate galaxy redshifts.

\section{Maximum-likelihood analysis}

\subsection{Overview}

We now have a model for the changes that a lens of given surface
density will produce in the $N(m,z)$ distribution of galaxies to a
given magnitude limit.  The next step is to design some procedure to
extract an estimate of the surface density from a given set of data,
and the obvious candidate is to use the likelihood methodology.

We begin by dividing the redshift and magnitude axes up into $q_1$ and
$q_2$ independent bins, each of which contains $n$ galaxies and has an
expected content of $\mu$ in the absence of lensing. The desired
likelihood function is then given by
\be
{\cal L}\propto\prod_1^q P[n|\mu(\kappa_\infty,z)].
\label{eq:likelee}
\ee
{}From this we can obtain an estimate of the lens strength by maximizing
${\cal L}$ with respect to $\kappa_{\infty}$, and `1$\sigma$' errors
on the estimate from
\be
\delta \kappa_\infty = \left\lgl
-\left({\partial^2 \ln {\cal L} \over \partial \kappa_\infty^2}
 \right)^{-1} \right\rgl^{1/2}.
\ee

The size of the bins is determined in order to fulfill the criteria of
statistical independence assumed in equation (31). In the case of
magnitude space, each galaxy is randomly selected from the luminosity
distribution, and hence completely independent. In this case the bins
can be made infinitely small, so that $q_2$ is the number of galaxies,
and $n=1$ per bin.

It is tempting to do this also in the case of the redshift
distribution. However, this would only be a good idea if the expected
distribution of background galaxies was Poissonian. In practice a
crucial limiting factor in this analysis will be galaxy clustering.
There is no point in making the bins shorter in radial extent than the
coherence length of galaxy clustering ($\simeq 10h^{-1}$ Mpc), as they
will then experience correlated fluctuations. It is therefore
important to calculate fully the probability distribution for $n$ in
the presence of both finite-$n$ fluctuations and fluctuations from
galaxy clustering; this is undertaken in the next section.

The problem of galaxy clustering in the redshift distribution raises
one further complication, which we mention here. Gravitational lensing
affects the background galaxies in two ways: it changes the shape of
the redshift probability distribution by reducing the fraction of
low-$z$ galaxies and boosting the proportion of high-$z$ galaxies. It
also produces a slight boost in total numbers to a fixed apparent
magnitude, since the latter effect generally overcomes the former. We
shall nevertheless neglect this effect in our analysis, and
concentrate only on the {\it shape\/} of the redshift distribution.
The reason for this is that in practice it is harder to obtain a
robust prediction of the background redshift distribution in absolute
terms: galaxy clustering in any normalization field off-cluster means
that the background surface density is not known precisely. If we
instead focus on the probability distribution for redshift, this
uncertainty is unimportant; this procedure also takes out part of the
effect of any galaxy clustering in galaxies behind the target cluster,
as well as making the calculation less sensitive to any uncertainty in
the exact limiting magnitude of the data.  In what follows, the
expected number of galaxies in a given bin, $\mu$, will therefore be
deduced by normalizing to the total observed number over the redshift
range over which the analysis is performed.

\subsection{Redshift  analysis}
\subsubsection{Effects of background galaxy clustering}

The problem to be solved is to find the probability distribution for
the number of galaxies in a given redshift bin, allowing for both
Poisson statistics and galaxy clustering.  As far as the latter is
concerned, it is relatively easy to calculate the fractional rms
number fluctuation, if we have some hypothesis for the clustering
power spectrum at the redshift of interest. If we call the power
spectrum $\Delta^2(k)$ (meaning power per log wavelength), then
the required rms $\sigma$ is just
\be
\sigma^2=\int \Delta^2(k)\; |W(k)|^2\; {dk\over k},
\ee
where $W(k)$ is the azimuthally-averaged Fourier transform of the
spatial bin under consideration.  One can similarly work out the
covariance between the numbers in different bins.  For power spectra
of interest, this turns out to be negligibly small for all but
adjacent bins.  For these, there is a small degree of coupling
(correlation coefficient $\simeq 0.2$), but we have neglected this and
treated the individual cells as independent. To within the accuracy to
which $\sigma^2$ can be calculated, this is an unimportant source of
error.

In principle, to work out the cosmic variance for a given power
spectrum and form of bin requires a 6-dimensional integral to be
performed numerically. Fortunately, things simplify a good deal in
cases of practical interest. The bin is defined by some angular
selection on the sky, and so its transverse extent is a function of
redshift. However, since we want reasonable redshift resolution,
$\delta z/z$ is small and it is a reasonable approximation to treat
the bin as having constant width. We then have a factorization into
the product of a radial window and a transverse window, and the
Fourier transform of the bin similarly factorizes into a product of
the two $k$-space windows.  We can further choose to make life easy by
picking windows with analytic Fourier transforms; if the angular
window is further chosen to be circularly symmetric, we are left with
only a two-dimensional integral over wavenumber $k$ and polar angle in
$k$ space.

For example, consider the case of a cylindrical bin of length $L$ and
radius $R$; the more practically interesting case of transverse
selection in some annulus can be obtained simply from this. The window
function is
\be
W(k,\mu)={\sin \mu kL/2\over\mu kL/2}\;
{2\over kR\,\sqrt{1-\mu^2}} J_1(kR\,\sqrt{1-\mu^2}),
\ee
where $\mu={\bf \hat k\cdot \hat r}$ is the cosine of the polar angle.
Given a power spectrum, we can now find the cosmic variance for any
given bin width and length. The simplest model which is realistic is
to use the Fourier transform of the canonical small-separation
correlation function, $\xi(r)= [r/5h^{-1}{\rm Mpc}]^{-1.8}$; the true
power spectrum curves below this function at large wavelength (Peacock
1991), so this is a conservative calculation. It will be important to
include evolution, since we expect that clustering at high $z$ will be
less than today. In general the shape of the power spectrum is
expected to alter as well as its amplitude, but we can only allow for
this by the rash step of picking a specific physical mechanism for the
evolution. We prefer the empirical approach of allowing for a scaling
of the clustering amplitude by some power of $(1+z)$. The model for
the power spectrum is thus
\be
\Delta^2(k)=0.903 [5 (k/h\,{\rm Mpc}^{-1})]^{1.8}\; (1+z)^{-\epsilon}.
\ee
We work throughout with comoving length units; $\epsilon=0$ thus
corresponds to the case of `painted-on' clustering that expands with
the Hubble flow; $\epsilon=2$ corresponds to linear-theory evolution,
and is close to what appears to be required by recent data on
faint-galaxy clustering ({\it e.g.} Couch, Jurcevic \& Boyle 1993).
Figure 3 shows the redshift dependence of the variance with this
evolution.

Of course, we need more than just a variance to specify the
distribution of galaxy counts. A useful model to adopt for this is the
Lognormal ({\it e.g.} Coles \& Jones 1991).  This not only modifies
the common assumption of a Gaussian random field to satisfy the
physical constraint of positivity, it also has some empirical support
going back to Hubble.
{}From the point of view of fitting the lens model, it is not
too critical that the lognormal model applies; the reason for
adopting it is that it provides a convenient means for
generating realistic mock datasets for testing our algorithms.
The way the lognormal model works is to
generate a Gaussian density fluctuation, $\delta$, of mean zero and
rms $\sigma$, and to construct a new density perturbation
\be
1+\delta'=\exp[\delta-\sigma^2/2].
\ee
The last term here is a normalization factor: $\langle \exp\delta
\rangle=\exp[\sigma^2/2]$.  Furthermore, the variance of the lognormal
density field as defined above is $\langle (\delta')^2\rangle
=\exp[\sigma^2]-1$; the parameter $\sigma^2$ should therefore be
\be
\sigma^2=\ln\left[1+\langle (\delta')^2\rangle\right],
\ee
where $\langle (\delta')^2\rangle$ is the cosmic
variance calculated from the observed power spectrum.

To incorporate finite-$N$ fluctuations, we assume that the observed
number of galaxies, $n$, is drawn Poissonianly from an expected
number, $\bar n$, which is itself subject to lognormal fluctuations
about some ensemble average $\mu$. The overall probability of getting
$n$ galaxies is then given in terms of the Poisson and Gaussian
distributions by
\be
P(n|\mu, \sigma)=
\int P_{\ss P}[n|\mu\exp(x)]\; dP_{\ss G}(x).
\label{eq: redlike}
\ee
The parameter $\mu$ in this equation is $\exp -\sigma^2/2$ times the
ensemble mean, for the reasons of normalization discussed above.


In Figure 4, we show a series of background redshift distributions
to demonstrate the
effect of varying the lens amplification and of the variance due to
lognormal density fluctuations. The lognormal field (filled dots) can
be seen to fluctuate increasingly about the underlying mean observed
density field (solid line) as the variance is increased (from left to
right).  Further scatter about the mean field is induced by the
Poissonian sampling of the lognormal field (histogram). Clearly, in
extreme cases of a highly evolved density field ($\sigma=1$), the
underlying mean field is difficult to recover.

\subsubsection{Maximum likelihood analysis of the redshift distribution}

Using this statistical procedure we can now construct a maximum
likelihood function, based on the random sampling of a lognormal
density field.  We divide the redshift distribution into bins larger
than the correlation length of clustering, and assume that each bin is
uncorrelated. Using the probability distribution of equation (38) for
the number of galaxies in a cell, the likelihood function over all
bins is
\be
{\cal L}(\kappa_{\infty}|n,\mu,\sigma)
= \prod_i P(n_i|\mu,\sigma_i),
\ee
where we have again parameterized the distortion in terms of the
surface density at infinity.

In Figure 5a,b we show the unlensed and lensed redshift distributions of a
Poisson sampled lognormal field of 300 galaxies. For the bin size of
$\Delta z= 0.05$ and radius $5'$ used here, we expect the variance to be
approximately $\sigma=0.2$ at  redshift of $0.5$ (Figure 3).
  We used a singular isothermal
cluster model placed at a redshift of $z=0.2$, and with a lens convergence
amplitude at infinity of ${\bar\kappa}_{\infty}=0.2$. This value for the mean
convergence interior to the limiting radius
corresponds to a 1D velocity dispersion $\sigma_v\simeq 1500\;\rm km\,s^{-1}$
for an isothermal sphere.

Figure 5c shows the normalized likelihood function as a function
of $\kappa_\infty$ for the reconstructed lens producing a maximum
likelihood value of $\kappa_{\infty}=0.2\pm 0.06$, which demonstrates that the
input amplification can be recovered. Extending the
model to include a homogeneous, negative contribution from cluster
dust, parameterized by $\kappa_{{\rm dust}}=-0.2$, the lensed and dust
obscured redshift distribution is shown in Figure 6a,b. Keeping the dust
obscuration as a free parameter, the calculated likelihood
contours are calculated and plotted in Figure 6c. The contours are separated
by $\Delta \ln {\cal L} = -0.5$. The likelihood function is maximized in
parameter space for values of $\kappa_{\infty}=0.2 \pm 0.07$ and
$|\kappa_{{\rm dust}}|=0.18 \pm 0.015$. Note we actually
have more information of the values of $\kappa_{\infty}$ and
$\kappa_{{\rm dust}}$, given that these must be, respectively,
 positive-definite and
negative-definite parameters. It is apparent from these
figures that the recovery of the amplification factor of the lens
along the redshift axis is viable, even with the inclusion of
obscuration by dust.

\subsection{Magnitude--space analysis}

As noted earlier, the statistical analysis in magnitude--space
is considerably easier than redshift--space due to the statistical
independence of galaxies. This follows directly from equation (1) in
the limit that galaxy luminosities are independent of environment.

The shift in apparent magnitude distribution at each redshift compared
with the field can be modelled by the conditional probability function
\be
p(m|\kappa_{\infty},z) = \frac{ N(m+2.5\log_{10}A(z), z)}{
\int N(m+2.5\log_{10}A(z), z)\; dm}
\ee
where we use the model Schechter function luminosity function (Section
2.4), normalized in each redshift bin to the mean field value.  Thus
fluctuations in the amplitude of the luminosity function due to
density perturbations are normalized away.

In this case, we can once again use a likelihood analysis, only this
time there is no objection to making the bins infinitesimally small.
Since the galaxy population randomly samples luminosity space, each
slice in redshift space is statistically independent. Using equation
(40) for the probability distribution of luminosities in a redshift
slice, the probability of each galaxy occurring in thin slice can be
calculated. Hence the likelihood of finding each galaxy in a small
volume about $(m_i,z_i)$ in redshift/magnitude space is
\be
{\cal L}(\kappa_{\infty}|m) = \prod_{i=1}^{n} p(m_i|\kappa_{\infty},z_i),
\ee
where the product is over all galaxies, $n$ is the total number of
galaxies, and again we have parameterized the amplification in terms
if the surface density at infinity.

Figure 7a,b shows two realizations of the distribution in
redshift/magnitude space. Figure 7a is the unlensed case, while
Figure 7b has the singular isothermal lens at $z_{\ss L}=0.2$, as before.
Again we use a lens magnification corresponding to
$\kappa_{\infty}=0.2$.  The brightening of galaxies beyond the lens is
clearly discernible, while the fluctuations in front of the lens are
due to shot noise and clustering.

Figure 7c shows the normalized likelihood function, which is maximized for
$\kappa_{\infty}=0.22 \pm 0.06$. While this is comparible with
 the redshift--distribution method for this
particular sample size, as we have already discussed and shall
show quantitatively in the next section, there is no lower bound on
the accuracy of this method caused by the intrinsic galaxy clustering.

Extending the model to include the homogeneous dust
distribution, with $\kappa_{{\rm dust}}=-0.2$ and no lensing, Figure
8a shows that the high redshift amplification is again
heavily suppressed. Figure 8b shows how the lens acts
against this obscuration. The likelihood contours
for this model are show in Figure 8c, where the spacing of contours is
$0.5$ in log likelihood. The outer contour corresponds to the
normalized likelihood of $e^{-5}$. Maximising the function gives us a
solution of $\kappa_{\infty}=0.21 \pm 0.05$ and $|\kappa_{{\rm
dust}}|=0.22 \pm 0.015$.

Given these results for the simulated lensing case, we now
proceed to compare the two methods as a function of sample size, and
discuss the robustness of our results.

\section{Practical application}

\subsection{Comparison of methods as a function of sample size}

In order to compare the two methods, we have run simulations
on the above lines, varying the interesting parameters of
lens redshift, depth and area on the sky.

For the redshift-distribution method, the error in $\kappa_{\infty}$
can be modelled by
\be
\sigma^2_{(z)}=
S_1^2(z_{\ss L},m_{\rm lim}) + S_2^2 (z_{\ss L},m_{\rm lim},\theta),
\ee
where
\ba
   S_1 &\simeq& {1.1+ z_{\ss L}10^{-0.2(m_{\rm lim}-22.5)}\over \sqrt{n}} \\
   S_2 &\simeq& 0.05\,10^{-0.56z_{\ss L}(m_{\rm lim}-19.5)}
\ea
Interestingly, $S_2$ turns out to be almost independent of $\theta$:
the pencil beams of practical relevance ($\theta\sim 10'$) are so thin as to be
dominated by structures much larger then their widths.

In the case of the magnitude estimator, the variance is invariant to
density perturbations and the error is purely shot--noise limited.
Again we can model this dependence by
\be
\sigma_{(m)} \simeq {0.8+ z_{\ss L}10^{-0.2(m_{\rm lim}-23.5)}\over \sqrt{n}}
\ee
In both cases, $n$ is the number of field galaxies
to the sample limit (including those in the foreground of the lens).
The observed number will of course be augmented by cluster
galaxies, but these would be removed in practice by ignoring
the redshift bins covering the cluster.
The magnitude limit, $m_{\rm lim}$ is for $R$ band, but other
wavebands could be used by scaling to limits with the same
median redshift.

We see that both methods are comparable, and that the redshift
estimator is the less accurate, except for small samples and
shallow limits.  This is reasonable, given that the area dilution
effect means that the redshift-estimator signal is largely
confined to the few very high-redshift galaxies.

As both approaches are statistically independent of one another, we
can simply multiply the two likelihood estimators
\be
{\cal L}(\kappa_{\infty}) = {\cal L}(\kappa_{\infty}|n,\mu,\sigma)
 {\cal L}(\kappa_{\infty}|m),
\ee
and achieve a total error of $[1/\sigma^2_{(z)}+1/\sigma^2_{(m)}]^{-1/2}$.
Clearly, however, it will be preferable to
make the two estimates independently and check them for agreement
before combining the methods.

The analysis given above assumes that we know the luminosity function
of the field population exactly. In practice, this must be estimated
from the data. To minimize systematics, the preferable procedure would
be to have equivalent data on a cluster and on a number of random
comparison fields.  We can simulate this procedure here by simulating
a field dataset without lens, using a Schechter function of known
parameters, fit a new Schechter function to the field realization, and
use this to analyze a lens simulation. Clearly, such a procedure will
introduce an additional error into any estimate of $\kappa_\infty$, of
the same order as that which applies if the luminosity function is
known exactly.  If several comparison fields exist, this should not be
an important source of error.

\subsection{Observing strategy}

Given the above results, we can ask what is the optimum
approach for detecting a given level of dark matter in the
minimum telescope time.
To see how this scales, we simplify the analysis by
assuming Euclidean space and the Poisson limit. This
means that the rms error in $\kappa(z_{\ss S})$ scales directly as
$n^{-1/2}$, for $n$ galaxies. In the low-redshift limit, this
says that the number of galaxies required to measure a given $\Sigma$ scales as
\be
n\propto {z_{\ss S}^2\over (z_{\ss S}-z_{\ss L})^2 z_{\ss L}^2}.
\ee
Now, since we are interested in a given area of the cluster,
the number of galaxies available is
\be
n_{\rm tot}\propto z_{\ss L}^{-2}z_{\ss S}^3.
\ee
The maximum signal-to-noise that can be obtained is thus independent
of lens redshift for $z_{\ss L}\ll z_{\ss S}$, but low lens redshifts
require the measurement of more redshifts by virtue of the larger
area of sky covered by the cluster.

We now assume that the spectroscopy is background limited,
so that the time taken to obtain a given
redshift scales as $({\rm flux})^{-2}\propto z_{\ss S}^4$, and
the total time is
\be
t\propto z_{\ss}^4 n\propto
{z_{\ss S}^6\over (z_{\ss S}-z_{\ss L})^2 z_{\ss L}^2}.
\ee
The optimum has $z_{\ss L}=z_{\ss S}/2$ and the total time
scales as $z_{\ss S}^2$. It is therefore much faster to
try to detect the effect by observing bright galaxies and
cluster lenses at low redshift. However, the total number
of available background galaxies will then be rather small, setting
an upper limit to the sensitivity of the measurement.
The way round this problem will be to stack clusters
statistically, obtaining an average dark matter profile.
Although maps for individual clusters would of course be
preferable, the increase in speed makes this an attractive
way of proceeding initially.

As a concrete example, consider the limit $R=20$ which is the
practical limit for fibre spectrographs; the surface density
here is about 1700 $\rm deg^{-2}$. Consider clusters at $z_{\ss L}=0.1$,
so that $1h^{-1}$ Mpc radius corresponds to $13.5'$ radius
and about 270 available galaxies. For a
$\sigma_v=1000 \;\rm km\,s^{-1}$ isothermal sphere, the integrated
surface density within this radius is ${\bar\kappa}_\infty=0.03$.
{}From the above calculations, the redshift and magnitude errors
on the $\kappa_\infty$ estimates would be $\sigma_{(z)}\simeq 0.086$
and $\sigma_{(m)}\simeq 0.079$, or a combined rms of 0.062.
The observation of 26 clusters would thus permit the
detection of this level of dark matter at the $2.5\sigma$ level.

\section{Limited redshift information}

\subsection{Colour--redshift estimates}

The need for redshift information for the above methods can be partly
overcome with colour information.  At the very least, one can strip
away the foreground, by identifying the subset of galaxies redder than
the cluster E/S0 galaxies.  In principle, given multiwavelength data,
it should be possible to estimate redshifts to some limited accuracy
for each galaxy.  Here one can achieve much fainter magnitudes and
hence higher surface densities, trading off the shot--noise
contamination against uncertainties in redshift estimation.

Many redshift--independent distance indicators have a power--law
relation to the distance, and so the associated errors in redshift may
be assumed to have a lognormal distribution.  The conditional
probability of finding a galaxy at redshift $z_e$ given that its true
redshift is $z$ is
\be
p(z_e|z) dz = \frac{dz}{\sqrt{2 \pi} \sigma_z z}
\exp \left( - \frac{1}{2 \sigma_z^2}(\ln z_e - \ln z)^2
\right).
\ee
The required probability distribution for the luminosity function is
then
\be
p(L|z_e) = \int  p(L|z) p(z|z_e) \, dz.
\ee

The likelihood functions for redshift errors of 5 and 10 $\%$ yield
uncertainties in $\kappa_{\infty}$ of $0.10$ and $\sim 0.4$ respectively.
Clearly errors in the redshift measurement greater than 5 $\%$ are unacceptable
for our purposes.

\subsection{Number--magnitude counts}

In the limit that we ignore redshift information altogether, the opportunity
to avoid clustering noise is lost. However, the shot--noise contamination is
now minimised due to the larger sample size expected from imaging galaxies
compared with spectroscopy. Again colour information can be used to remove the
cluster E/S0 sequences.

The count slope will be flattened for a fixed amplification since the slope of
the magnitude distribution for the field at faint magnitudes is a decreasing
function of magnitude, particularly in the near IR where the K-correction
dominates over the evolution. This flattening of the count slope is more
interesting than the change in amplitude of the counts since it is less
susceptible to the clustering fluctuations than the total number.

We may calculate the feasibility of the detection of such a slope change in
the background counts by simply projecting our reconstructions into 2--D. Using
the usual isothermal lens model at a redshift of $z_{\ss L}=0.2$, Figure 9
shows the theoretical number--magnitude distribution radially interior to
mean convergences of $\kappa_{\infty}=
0, \, 0.1, \, 0.5$ and $1$ respectively. The
main feature is the expected enhanced curvature of the distribution.
Superimposed on the plot is the unlensed case of a random selection of $3000$
galaxies, in the presence of the same degree of clustering
as used above.

We estimate that the error introduced here is of order 0.2 in
$\kappa_\infty$, so this approach is of limited accuracy
Furthermore, the required
number of galaxies will only be available over areas where $\kappa_\infty$
is low ($\kappa_\infty\ls 0.05$ for 3000 galaxies to $R=25.5$ in
this case).
Nevertheless, such a method may be useful for statistical
averaging over a number of clusters.

\subsection{Size shifts}

In principle, the most promising method may be one
that uses the lens signature directly. If we plot the
field-galaxy population as some measure of galaxy size
versus surface brightness, then the effect of a lens
is to move all galaxies to larger size at constant
surface brightness. The fractional increase in size
is just $\kappa(z_{\ss S})$, which will typically be
a few \% for the examples illustrated earlier.
Given sufficient background galaxies, it should be
possible to detect this shift. The practicalities involve
the effects of seeing, but we note that
the reduction in sensitivity to size shifts caused by
seeing will be similar to the reduction in sensitivity
to shear. For data with $0.5''$ FWHM seeing, Fahlman \et (1994)
demonstrated only a 30\% reduction in shear sensitivity, and
obtained rms shear limits of $\sim 1\%$ by averaging over
about 2500 galaxies.
Provided any effect due to cluster galaxies can be
removed (by using a colour criterion), this paints
an encouraging prospect for the direct detection of lensing
via shifts in galaxy size.
We hope to investigate this method in detail elsewhere.

\section{Conclusions}

In this paper, we have presented a new method for measuring the mass
distribution in the outer parts of galaxy cluster haloes via
gravitational lensing.  This is certainly one of the most interesting
questions in cosmology: if a critical-density universe is to be
compatible with observations, clusters should have extended
quasi-isothermal dark haloes, and gravitational lensing
is probably the only method by which these can be detected
directly. We have tested our methods
on realistic simulated data, in the presence of shot noise
and fluctuations from galaxy clustering. We estimate that
our methods can realistically expect
to detect any dark matter halo around clusters out to
radii of at least $1h^{-1}$ Mpc.

Our reconstruction method is limited by finite numbers of
galaxies behind the cluster lens:
we must average over some area in order to gather sufficient
galaxies to define the redshift distribution.
To obtain a map of surface density to a given fractional
accuracy requires a number of galaxies per pixel which
scales roughly as $r^2$ for an isothermal halo.
Thus, although in principle one might be able
to obtain a genuine map of density with arcmin resolution
of the central parts of a cluster, in the most interesting
outer regions the method can give only a radial profile in
a set of increasingly coarse annuli.
Nevertheless, such information is quite adequate for answering
the critical questions concerning the relative distributions
of mass and light. If we are confined to a radial profile,
there is no reason not to stack the signal from several
clusters in order to obtain a statistical detection of
dark matter at large radii. We have shown that such an experiment
can be made relatively economical in terms of telescope time.

Finally, it is interesting to compare our method with that of Kaiser
\& Squires. Each technique has strengths and weaknesses, which are
largely complementary.  The virtue of our method is that it measures
the surface density directly, rather than effectively having to
differentiate the (noisy) shear.
Moreover, we thus avoid the principal drawback of the Kaiser \&
Squires method, which is its insensitivity to a constant-density
screen and the corresponding need to assume $\Sigma=0$ at the data
boundary.  Lastly, the signature of lensing in our method is
uncomplicated (extra high-$z$ galaxies) and not vulnerable to subtle
systematics in the data.  The principal limitations of our method are
the sensitivity to galaxy clustering, plus the fact that spectroscopic
data are more time-consuming to obtain than deep imaging.
However, we believe we have demonstrated that our method
can be made to work well with datasets of a practical size and quality.
The very different approaches of the two reconstruction
methods is a great virtue: any case for which these techniques yield
concordant results deserves to be treated with a high degree of
confidence.

\bigskip
\ref\strut

\noindent
A.N.T. is supported by a SERC research assistantship.  We thank
Richard Bower, Richard Ellis, Pippa Goldschmidt, Alan Heavens, Ian
Smail and Alex Szalay for useful discussions.

\bigskip
\noindent{\bf REFERENCES}
\ref \strut
\ref Bonnet, H., Mellier, Y., Fort, B., 1994, \apj, in press.
\ref Bower, R.G, Lucey, J.R., Ellis, R.S., 1992, \mn, {\bf 254}, 601.
\ref Boyle, B.J., Fong, R., Shanks, T., 1988, \mn, {\bf 231}, 897.
\ref Broadhurst, T.J., Ellis, R.S., Glazebrook, K., 1992,
{\it Nature}, {\bf 355}, 55.
\ref Colless, M., Ellis, R.S., Broadhurst, T.J., Taylor, K.,
Peterson, B.A.,  1993, \mn, {\bf 261}, 19.
\ref Coles, P., Jones, B.J.T., 1991, \mn, {\bf 248}, 1.
\ref Couch, W.J., Jurcevic, J.S., Boyle, B.J., 1993,
\mn, {\bf 260}, 241.
\ref Dabrowski, M., Stelmach, J., 1986, AJ, {\bf 92}, 1272.
\ref Dekel, A., 1994, ARAA, in press.
\ref Fahlman, G., Kaiser, N., Squires, G., Woods, D., 1994, \apj, in press.
\ref Ferguson, H.C., 1993, \mn, {\bf 263}, 343.
\ref Glazebrook, K., Ellis, R.S., Broadhurst, T.J.,
1994, \mn, in press.
\ref Grossman, S.A., Narayan, R., 1989, \apj, {\bf 344}, 637.
\ref Kneib, J.-P., Mathez, G., Fort, B., Mellier, Y.,
Soucail, G., Longaretti, P.-Y., 1994, A\&A, in press.
\ref Kochanek, C.S., 1990, \mn, {\bf 247}, 135.
\ref Kaiser, N., Squires, G., 1993. \apj, {\bf 404}, 441.
\ref Metcalf, N., Shanks, T., Fong, R., Roche, N., 1994, \mn, in press.
\ref Miralda-Escud\'e, J., 1991, \apj, {\bf 370}, 1.
\ref Peacock, J.A. 1991, MNRAS, 253, 1p.
\ref Press, W.H., Gunn, J.E., 1973, \apj, {\bf 185}, 397.
\ref Refsdal, S., 1966, \mn, {\bf 132}, 101.
\ref Smail, I., Ellis, R.S., Aragon-Salamanca, A.,
     Soucail, G., Mellier, Y., Giraud, E., 1993,
     \mn, {\bf 263}, 628.
\ref Smail, I., Ellis, R.S., Fitchett, M.J., Edge, A.S., 1994,
     \mn, in press.
\ref Tyson, J.A., Valdes, F. Jarvis, J.F., Mills A.P.,
    1984, \apj, {\bf 281}, L59.
\ref Tyson, J.A., Valdes, F., Wenk, R.A., 1990.
 \apj, {\bf 349}, L1.
\ref White, S.D.M., Briel, U.G., Henry, J.P., 1993, \mn, {\bf 261}, L8.
\ref Young, P. 1981. \apj, {\bf  244}, 756.

\bigskip

\bfig[p]
{\small\caption{Series of graphs showing the distortion of a model
galaxy redshift distribution (from section 2.4) with increasing lens
amplification, $A$, according to equation (6).  The amplification
factor increases from 1 to 2. The effective local index of the
luminosity function, $\beta(z)$ is parameterized in Section 2.4.
Because this is a normalized probability distribution, the
distribution changes slightly at $z<z_{\ss L}$, even though
lensing clearly does not affect the number of galaxies there.
\label{fig: fig1}}}
\efig

\bfig[p]
{\small\caption{ The lens convergence amplitude,
$\kappa(z)/\kappa_{\infty}$, as a function of density
parameter, $\Omega_0$, in the range 0.1 to 1
and redshift of the lens, $z_{\ss L} = 0.1$ to $0.4$ (a)
showing the insensitivity to $\Omega_0$ for a given lens redshift.
Also (b) the dependence of the lens convergence amplitude,
$\kappa(z)/\kappa_{\infty}$, as a function of $\Omega_0$ and lens redshift,
$z_{\ss L}$, in a
flat universe ($\Omega_0 + \Omega_{\Lambda}=1$). $\Omega_0$ is in the
range 0.1 to 1 and $z_{\ss L}=0.1$ to $0.4$.
 Again, the ratio is highly insensitive to $\Omega_{\Lambda}$ for given lens
redshift.
\label{fig: fig3} }}
\efig

\bfig[p]
{\small\caption{ Plots of fractional density variance expected due to
galaxy clustering in circular bins of angular
radius $5'$ and length $\Delta z=0.05$,
as a function of redshift. Clustering is assumed to evolve at the
linear-theory rate ($\epsilon=2$).
\label{fig: fig4} }}
\efig

\bfig[p]
{\small\caption{ Lensed redshift distributions for
varying fractional density rms $\sigma$
and lens strength $\kappa_{\infty}$, assuming
$z_{\ss L}=0.2$ and a magnitude limit $R=22.5$. The underlying distribution
function (solid line), normalised to a sample of 300 galaxies,
is used to construct a lognormal density field (large dotted line),
which is then Poisson sampled (histogram). The top row has
$\kappa_{\infty}= 0.1$, while the bottom row has $\kappa_{\infty}=0.5$
for an isothermal lens.  The three columns have $\sigma=$0.1, 0.5, and
1, respectively.
\label{fig: fig5} }}
\efig

\bfig[p]
{\small\caption{ Unlensed (a) and lensed redshift distributions (b) for a
compounded lognormal--Poisson distribution of galaxies. The underlying
rms of density fluctuations is set at $\sigma=0.5$, and there are
a total of 300 galaxies in the simulation, corresponding
to bins of angular size $5'$ and length $\Delta z = 0.05$.
The convergence factor used is $\kappa_{\infty}=0.2$, for a singular
isothermal cluster.
The corresponding likelihood curve for the lensed distribution
is shown (c). The amplification is recovered with an rms
error in $\kappa_{\infty}$ of $\sigma_{\kappa_{\infty}}=0.06$.
\label{fig: fig6} }}
\efig

\bfig[p]
{\small\caption{ As for Figure 5, but with isothermal cluster model
extended to include a homogeneous distribution of dust. The dust acts
so as to suppress the effects of lensing. We parameterize the
suppression by $\kappa_{{\rm dust}}=-0.2$.
 The likelihood corresponding likelihood contours are plotted (c) with
 a spacing of $\Delta \ln {\cal L} =
-0.5$. The function is maximized by the parameters $\kappa_{\infty}=0.2
\pm 0.07$ and $\kappa_{{\rm dust}}=-0.18 \pm 0.015$.
\label{fig: fig7} }}
\efig

\bfig[p]
{\small\caption{ Magnitude/Redshift--space distortions. Figure 7a
is the unlensed distribution, while Figure 7b is lensed by a
constant-density screen. The lens is placed at $z_{\ss L}=0.2$, and the
lens amplification is parameterized by $\kappa_{\infty}=0.2$.
 The Likelihood function for magnitude--space
distortions is shown (c) for this case.  Again we find an rms
$\kappa_{\infty}=0.22\pm 0.06$, compared with the input value of 0.2.
\label{fig: fig8} }}
\efig

\bfig[p]
{\small\caption{ Magnitude/Redshift--space distortions. As in Figure
7a,b but with a homogeneous distribution of dust, uniformly
suppressing the lensing amplification beyond the lens.  Figure 8a
shows the distortion from dust only, while 8b is both dust and
lensing.  The negative contribution to the amplification by the dust
is parameterized by $\kappa_{{\rm dust}}=-0.2$.
Figure 8c shows likelihood contours for magnitude--space distortions
from lensing and dust obscuration.  Again for a sample of 300
galaxies, we find $\kappa_{\infty}=0.21 \pm 0.05$ and
$|\kappa_{{\rm dust}}|=0.22 \pm 0.015$.
\label{fig: fig9} }}
\efig

\bfig[p]
{\small\caption{ Number--magnitude distribution (normalised).
Theoretical curves for an isothermal lens at $z_{\ss L}=0.2$ with
$\kappa_{\infty}=0,0.1,0.5,1$ are shown. Superimposed is the
unlensed case for a sample of 3000 galaxies randomly sampled
from a lognormal density field with $\sigma=1$ at zero redshift.
The error on the corresponding estimate of
$\kappa_{\infty}$ would be approximately 0.2.
\label{fig: fig10} }}
\efig

\end{document}